\documentclass[aps,superscriptaddress,twocolumn,showpacs,floatfix,notitlepage]{revtex4-2}

\usepackage{amsmath}
\usepackage[makeroom]{cancel}
\usepackage{amsfonts}
\usepackage{amssymb}
\usepackage{bm}
\usepackage{graphicx}
\usepackage{color}
\usepackage{xcolor}
\usepackage{xspace}
\usepackage[sort&compress]{natbib}
\usepackage{dsfont}
\usepackage{physics}
\usepackage{siunitx}
\usepackage{xfrac}
\usepackage[cal=boondox,scr=boondoxo]{mathalfa}

\DeclareFontFamily{OT1}{pzc}{}
\DeclareFontShape{OT1}{pzc}{m}{it}%
{<-> s * [1.15] pzcmi7t}{}
\DeclareMathAlphabet{\mathpzc}{OT1}{pzc}{m}{it}

\newcommand{\la}{\circlearrowleft}
\newcommand{\ra}{\circlearrowright}

\newcommand{\ud}[1]{{#1^{\dagger}}}

\newcommand{\av}[1]{\langle  #1 \rangle}

\usepackage[colorlinks=true, linkcolor=blue, citecolor=magenta]{hyperref}

\allowdisplaybreaks

\begin{document}
\title{Spatial correlations of vortex quantum states}

\author{Eduardo Zubizarreta Casalengua}
\email{eduardo.zubizarretacasalengua@wsi.tum.de}
\affiliation{Walter Schottky Institute, School of Computation, Information and Technology and MCQST, Technische Universit\"at M\"unchen, 85748 Garching, Germany}
\author{Fabrice P.~Laussy}
\affiliation{Instituto de Ciencia de Materiales de Madrid ICMM-CSIC, 28049 Madrid, Spain}
\affiliation{Faculty of Science and Engineering, University of Wolverhampton, Wulfruna St, Wolverhampton WV1 1LY, UK}

\email{fabrice.laussy@gmail.com}

\date{\today}

\begin{abstract}
  We study spatial correlations of vortices in different quantum states or with Bose or Fermi statistics. This is relevant for both optical vortices and condensed-matter ones such as microcavity polaritons, or any platform that can prepare and image fields in space at the few-particle level.  While we focus on this particular case for illustration of the formalism, we already reveal unexpected features of spatial condensation whereby bosons exhibit a bimodal distribution of their distances which places them farther apart than fermions in over 40\% of the cases, or on the opposite conceal spatial correlations to behave like coherent states.  Such experiments upgrade in the laboratory successful techniques in uncontrolled extreme environments (stars and nuclei).
\end{abstract}
\maketitle

In a statistical theory, when waves interfere, they produce
correlations between them. This
elementary fact surprisingly escaped notice for centuries despite the
unanimous embrace and resounding successes of both wave theory and of
statistics. This changed when a young engineer (Hanbury
Brown)~\cite{hanburybrown_book91a} working on the secret development
of the radar for British intelligence, spotted the effect with his
naked eyes on oscilloscope traces and used it to create a new type of
stellar
interferometry~\cite{hanburybrown56a}. Purcell~\cite{purcell56a}
understood that this counter-intuitive effect---correlations from
non-interacting, indeed merely interfering objects---was in fact fully
expected from even more fundamental quantum mechanical considerations
and extended it to fermions, predicting instead anticorrelations as
opposed to bosonic positive correlations. The effect became a powerful
tool to measure sizes ranging from the diameters of stars
($\SI{e12}{\centi\meter}$) to high-energy nuclear matter
($\SI{e-12}{\centi\meter}$)~\cite{baym98a}. While in radioastronomy,
intensity interferometry has been described by Hanbury Brown as
``building a steam roller to crack a
nut''~\cite{hanburybrown_book91a}, the technique became indispensable
in high-energy physics and nuclear physics~\cite{weiner_book00a}. It
allowed the first measurement of the radius of interactions for the
proton-antiproton process from the amount of correlations of identical
pions formed in the nuclear reaction~\cite{goldhaber60a}. Since then,
it became a central, sometimes only, way to measure sizes and
lifetimes of high-energy particles and nuclear
reactions~\cite{boal90a}, being instrumental for instance in the
investigation of such extreme forms of matter as the quark-gluon
plasma formed in ultrarelativistic nucleus-nucleus
collisions~\cite{alice11a}.  In quantum optics, while the emphasis has
been on temporal correlations, technological progress, e.g., with
superconducting nanowire single-photon detectors~\cite{natarajan12a},
can capture increasingly high-resolution images from single-photon
pixelated detectors~\cite{wang23a}, thus bringing back spatial quantum
correlations to the fore of solid-state and condensed-matter
physics~\cite{lubin22a}.  In high-energy and nuclear physics, bosonic
correlations---bunching of events---are mostly investigated, since
fermionic correlations are shadowed by stronger particle interactions
(typically, Coulomb repulsion from charged particles). In contrast,
low-energy quantum optics became able to control the statistics of
photons through their underlying quantum state. Indeed, Glauber
understood coherence as the faculty of photons to neutralize their
intrinsic Boson correlations~\cite{glauber63a} to be detected as an
uncorrelated (Poisson) stream, which could be understood as the
wave-like limit of light where the particle aspect vanishes.  At the
other extreme, photons can also be made to behave like fermions by
exhibiting
antibunching~\cite{kimble77a,cohentannoudji79a,khalid24a}. The ability
to control, or at least characterize, quantum states from photon
correlations resolved spatially could endow spectroscopic techniques
with new and considerably enhanced opportunities as compared to their
Bose-Einstein interferometry counterpart~\cite{weiner_book00a}.

Here, we describe correlations of multi-particle quantum states
extended in real space (or other related spaces through the
appropriate transforms). To keep the discussion simple and to the
point, we focus on a particular case that interfaces between
optics~\cite{dennis09a, shen19a} and condensed matter, namely,
vortices~\cite{quinteirorosen22a}.  Microcavity
polaritons~\cite{kavokin_book17a}, sitting halfway between optics and
the solid state, and with a particularly fruitful development of
vortices~\cite{lagoudakis08a,fraser09a}, provide but one example of a
platform where to pursue the next generation of bosonic
interferometry. As the topological cornerstone of 2D quantum fields,
vortices are central to many polaritonic
breakthroughs~\cite{sanvitto10b,krizhanovskii10a}, from the
observation of their BKT phase transitions~\cite{caputo18a} to, more
recently, the realization of the superfluid bucket
experiment~\cite{gnusov23a,delvalleinclanredondo23a} passing by
Rabi-propelled dynamics~\cite{dominici23a} or quantum
turbulence~\cite{panico23a}.  As quasi-particles which admix both
light (photons) and matter (semiconductor electron-hole pairs, or
excitons), polaritons have always elicited a quantum-mechanical
formulation, in particular as superpositions with a wavefunction
$\ket{\psi}\equiv\alpha(t)\ket{1_a,0_b}+\beta(t)\ket{0_a,1_b}$ that
entangles a photon (with Bose creation operator~$\ud{a}$) and the
exciton vacuum to its Rabi-flop counterpart with one exciton
(operator~$\ud{b}$) and no photon. This genuinely quantum-mechanical
state was, however, not realized until recently, by direct excitation
of a microcavity with quantum
light~\cite{cuevas18a,suarezforrero20a}. In all other cases, a product
state $\ket{\alpha(t)}_a\otimes\ket{\beta(t)}_b$ (no entanglement) or
something similar (with some amount of squeezing~\cite{boulier14a,
  delteil19a,munozmatutano19a} but still within Gaussian states), is
realized instead. In such cases, the probability amplitudes
of~$\ket{\psi}$ become classical-field amplitudes for coherent states
of both the photon and exciton
fields~\cite{dominici14a}. Because~$\alpha(t)$ and $\beta(t)$ follow
the same equations of motion in both interpretations, although these
are very different (quantum probabilities or classical-field
amplitudes, respectively), there tends to be some confusion as to the
exact quantumness involved with polaritons. One of the outcomes of
studying spatial correlations of photons emitted by the polariton
field will be to provide clear-cut experimental resolves of the
central and long-standing question of their quantum character.  We
also consider fermionic statistics, that can be realized for spatial
wavefunction with singlet-spin.  Our considerations are not specific
neither to polaritons nor to vortices, which merely provide an
illustration of the potential of spatial multiphoton correlations.

Quantum mechanics is a wave theory, and at the heart of its formalism
lie the eigenmodes of the system being described. Vortices provide a
textbook case of basis states for the 2D harmonic oscillator, for
which they provide eigenstates of defined angular momentum in terms of
the 1D eigenfunctions
$\phi_n (x) \equiv \big(\sqrt{2^n
  n!}\pi^{1\over4}\big)^{-1}e^{-{x^2/2}} H_n(x)$ with
$H_n(x)$ the Hermite polynomials with $n\ge 0$ integers
and~$x$ in units of the single-charge vortex radius.  It will be
enough for our discussion to consider stationary 2D vortices with
$\pm1$ topological charge, denoted $\ket{\la}$
and~$\ket{\ra}$, with eigenfunctions
$\phi_\circlearrowleft(\mathbf{r})=\phi_\circlearrowright^*(\mathbf{r})\equiv\braket{x,y}{\circlearrowleft}=\big[\phi_1(x)\phi_0(y)+i\phi_0(x)\phi_1(y)\big]/\sqrt2$. Given
the separation of variables and linearity of quantum mechanics, we can
actually develop the formalism for one variable only and keep our
discussion one-dimensional. This will be at no loss of generality but
with great simplifications of the notations.
In the second quantization formalism, the family of
$n$-particle density distributions is written in terms of a field
operator~$\hat\Psi^{\dagger}(x)\equiv\sum_m \phi_m^{*}(x)
a_m^{\dag}$ where
$\phi_m(x)$ is the one-particle wavefunction corresponding to the
$m$-th spatial mode $\ket{m}$ with creation operator
$a_m^{\dagger}$ (so
$\braket{x}{m}=\phi_m(x)$). In 2D, here we would have to keep track
of, say, $n_\la$
and~$n_\ra$ the indices labeling the basis. This operator applied on
the vacuum creates a particle at the position~$x$
since 
$\hat\Psi^{\dagger}(x)\ket{0}=\ket{x}$~\cite{sup} and yields
the density operator $\hat{n}^{(1)}(x) \equiv
\hat\Psi^\dagger(x) \hat\Psi
(x)$.  We now consider quantum states for the particles. The most
general case is described by the density matrix $
\hat\rho\equiv\sum_{\mathbf{m},\mathbf{n}}\alpha_{\mathbf{n}\atop\mathbf{m}}\ket{\mathbf{n}}
\bra{\mathbf{m}}$
%
%
where~$\mathbf{n}$,
$\mathbf{m}$ are vectors of integers that specify how many modes are
occupied in the occupation-number formalism with normalization
$\sum_{\mathbf{m},\mathbf{n}}|\alpha_{\mathbf{n}\atop\mathbf{m}}|^2=1$,
e.g.,
$\ket{\mathbf{n}}=\ket{n_0,n_1,\cdots,n_m}$ refers to the state
with~$n_0$ particles in the eigenstate~$\phi_0$,
$n_1$ in the
eigenstate~$\phi_1$, etc. Typically, one studies one-particle
observables, such as the density profile (e.g., near-field imaging),
recovered from the full quantum picture with the so-called reduced
one-particle density matrix $\rho^{(1)} (x) \equiv \langle
\hat{n}^{(1)} (x) \rangle = \int \bra{\mathbf{x}} \hat\rho \,
\hat{n}^{(1)}(x) \ket{\mathbf{x}}d\mathbf{x}$
where~$\ket{\mathbf{x}}$ is the vector with as many variables as
necessary to dot the operator~$\hat\rho\,\hat
n_1(x)$ which, since it can have a varying number of particles, makes
the integral running over possible particle numbers.  It reads in
terms of the single-particle wavefunctions and the quantum
state~$\hat\rho$~\cite{sup}:
\begin{equation}
  \label{eq:Wed27Sep165043BST2023}
  \rho^{(1)} (x) = \sum_{p,q} \av{a_p^{\dagger} a_q} \phi_p^* (x) \phi_q (x)
\end{equation}
where
$\av{a_p^{\dagger} a_q}\equiv\tr{\hat\rho a_p^{\dagger}
  a_q}=\sum_{\mathbf{m},\mathbf{n}}\alpha_{\mathbf{n}\atop\mathbf{m}}\bra{\mathbf{m}|\ud{a_p}a_q}\ket{\mathbf{n}}$. Similarly,
the reduced two-particle density
matrix~$\rho^{(2)} (x,x') = \langle{:}\hat{n}^{(1)} (x)
\hat{n}^{(1)}(x'){:}\rangle$
is obtained as~\cite{sup}:
\begin{equation}
  \label{eq:Wed27Sep172401BST2023}
  \rho^{(2)} (x,x') = \sum_{p,p',q,q'} \av{a_p^{\dagger}  a_{p'}^{\dagger} a_{q'}  a_q} \phi_p^* (x) \phi_{p'}^* (x')\phi_q(x)\phi_{q'}(x')\,.
\end{equation}
The integration of $\rho^{(1)} (x)$ and $\rho^{(2)} (x,x')$ over
$x$ and $x'$ leads to $\av{\hat{N}}$ and
$\av{{:}\hat{N}^2{:}}$, respectively, where~$\hat
N\equiv\sum_m\ud{a_m}a_m$. Therefore, they do not correspond to a
probability distribution but they can be normalised to act as such.
One could carry on with higher-particle numbers but it will be enough
for our present discussion to limit to two particles. The second
quantization formalism conveniently embeds self-consistently the
quantum statistics (or lack thereof) through the algebra of the
operators~\cite{sup}, so one merely has to compute $\av{a_p^{\dagger}
  a_q}$ and $\av{a_p^{\dagger} a_{p'}^{\dagger} a_{q'}
  a_q}$ in the chosen basis for the states considered.  When $p =
p'$ or $q =
q'$, the fermionic correlators give zero while the bosonic ones get
magnified by a factor
$\sqrt{2}$.  We are now in a position to compute the one- and
two-particle reduced density matrices for cases of interest. At this
stage, we can upgrade the formalism to the required dimensionality
through the
substitution~$x\to\mathbf{r}$ and the corresponding labeling of the
basis states. In our chosen case of single-charge vortices, i.e.,
remaining within a closed set of two states, the calculations are
straightforward.  We provide the general result for any two
modes~$a$
and~$b$ for the one-particle reduced density matrix in
Eqs.~(\ref{eq:Wed27Sep151331BST2023}) and turn to the particular case
of vortices for the two-body reduced density matrices in
Eqs.~(\ref{eq:Wed27Sep151754BST2023}).  We consider i) Fock states,
i.e., with an exact number of quanta, of both Bosonic and Fermionic
particles, so that in our two-modes picture, that can only
be~$\ket{1_a}_\mathrm{F}\ket{1_b}_\mathrm{F}$ for fermions while one
can distribute~$n$ bosons into $n-k$ in one mode
and~$k$ in the other:
$\ket{(n-k)_a}_\mathrm{B}\ket{k_a}_\mathrm{B}$. Now considering bosons
only, we can also turn to other quantum states, e.g., coherent
states~$\ket{\alpha}_a\ket{\alpha}_b$
where~$\alpha\in\mathbb{C}$ and $\ket{\alpha}\equiv
e^{-|\alpha|^2/2}\sum_{n=0}^\infty{\alpha^n\over\sqrt{n!}}\ket{n}$ or
thermal
states~$\rho_\theta\equiv(1-\theta)\sum_{n=0}^\infty\theta^n\ketbra{n}{n}$
with~$\theta\equiv{\bar n\over1+\bar
  n}$ the effective temperature for a thermal state with mean
occupation~$\bar n$.  So-called cothermal
states 
provide useful interpolations as mixtures of a coherent state of
intensity~$|\alpha_c|^2$ with a thermal state of
temperature~$\theta_c$. These are discussed in details in the
Supplementary~\cite{sup}. Obviously, numerous other cases (e.g.,
squeezing) could also be usefully added.

\begin{widetext}
  One-particle density matrices for quantum states of interest
  distributed over two modes~$\phi_a$ and~$\phi_b$ of the system:
  \begin{subequations}
    \label{eq:Wed27Sep151331BST2023}
    \begin{align}
      \text{Fock Fermions}~\ket{1_a}_\mathrm{F}\ket{1_b}_\mathrm{F}&:& \rho^{(1)}_\mathrm{F}(\mathbf{r}) &= |\phi_a (\mathbf{r})|^2 + |\phi_b (\mathbf{r})|^2 \,,\label{eq:Wed27Sep181102BST2023}\\
      \text{Fock Bosons}~\ket{n_a}_\mathrm{B}\ket{m_b}_\mathrm{B}&:&\rho^{(1)}_\mathrm{B}(\mathbf{r}) &= n |\phi_a (\mathbf{r})|^2 + m \, |\phi_b (\mathbf{r})|^2 \,,\label{eq:Wed27Sep181029BST2023}\\
      \text{Coherent state}~\ket{\alpha_a}\ket{\alpha_b}&:&\rho_{\mathrm{coh}}^{(1)}(\mathbf{r}) &= \big\vert \alpha_a \phi_a (\mathbf{r})+\alpha_b\phi_b(\mathbf{r}) \big\vert^2 \,,\label{eq:Wed27Sep181331BST2023}\\
      \text{Thermal state}~\rho_{\theta_a}\rho_{\theta_b}&:&\rho_\mathrm{th}^{(1)}(\mathbf{r}) &= \bar{n}_a |\phi_a(\mathbf{r})|^2 +\bar{n}_b |\phi_b(\mathbf{r})|^2\,,\label{eq:Wed27Sep181401BST2023}
  \end{align}
\end{subequations}
The corresponding two-particle reduced wavefunctions for vortices:
\begin{subequations}
  \label{eq:Wed27Sep151754BST2023}
  \begin{align}
    \text{Fock Fermions}~\ket{1_\la}_\mathrm{F}\ket{1_\ra}_\mathrm{F}&:&\rho^{(2)}_\mathrm{F}(\mathbf{r},\mathbf{r}')&=|\phi_\la (\mathbf{r}) \phi_\la (\mathbf{r}')- \phi_\ra (\mathbf{r}) \phi_\ra (\mathbf{r}')|^2\,,\\
    \text{Fock Bosons}~\ket{n_\la}_\mathrm{B}\ket{m_\ra}_\mathrm{B}&:&\rho^{(2)}_\mathrm{B} (\mathbf{r},\mathbf{r}') &= nm \, |\phi_\la (\mathbf{r}) \phi_\la (\mathbf{r}')+\phi_\ra (\mathbf{r}) \phi_\ra (\mathbf{r}')|^2 + {} \nonumber \\
                                                                     &&& {} + n(n-1) |\phi_\la(\mathbf{r})\phi_\la(\mathbf{r}')|^2 + m(m-1) |\phi_\ra(\mathbf{r})\phi_\ra(\mathbf{r}')|^2\\ 
    \text{Coherent state}~\ket{\alpha_\to}\ket{\alpha_\uparrow}&:&\rho_{\mathrm{coh}}^{(2)} (\mathbf{r},\mathbf{r}') &= \rho_{\to}^{(1)}(\mathbf{r})\rho_{\uparrow}^{(1)}(\mathbf{r}') \,,\label{eq:Tue16Jan194447CET2024}\\
    \text{Thermal state}~\rho_{\theta_\la}\rho_{\theta_\ra}&:&\rho^{(2)}_\mathrm{th}(\mathbf{r},\mathbf{r}') &= \sum_{p,p'\in\{\la,\ra\}} \bar{n}_p \bar{n}_{p'} \big(|\phi_p(\mathbf{r})|^2 |\phi_{p'}(\mathbf{r}')|^2 + 
                                                                                             \phi^*_p (\mathbf{r}) \phi_p (\mathbf{r}') \phi^*_{p'} (\mathbf{r}') \phi_{p'} (\mathbf{r})	\big) \,,
  \end{align}
\end{subequations}
\end{widetext}
It was necessary to consider the general case of two modes~$a$
and~$b$ in Eqs.~(\ref{eq:Wed27Sep151754BST2023}) because for the
configuration that interests us,
where~$\phi_a=\phi_\la$
and~$\phi_b=\phi_\ra$, which is such that
$|\phi_\la|^2=|\phi_\ra|^2$, then for conditions that make those
various states comparable, e.g., with the same mean populations, they
are exactly identical at the one-particle level, i.e.,
$\rho^{(1)}_\mathrm{F}=\rho^{(1)}_\mathrm{B}=\rho^{(1)}_\mathrm{coh}=\rho^{(1)}_\mathrm{th}=\rho^{(1)}_\mathrm{coth}=\rho^{(1)}_\mathrm{cat}$
wih also cothermal and cat states from the Supplementary~\cite{sup}
and still other quantum states could be added to this list. The
required adjustment
are~$n_\la=m_\ra=1$ for bosons (one in each state, like Fermions, thus
differing only in their statistics) and $\bar n_\la=\bar
n_\ra=1$ for thermal states, i.e., with same mean but thermal
fluctuations. For coherent states, besides the same mean, one further
needs to turn to the cartesian basis of dipoles,
$\phi_\to(\mathbf{r})\equiv\phi_1(x)\phi_0(y)$ and
$\phi_\uparrow(\mathbf{r})\equiv\phi_0(x)\phi_1(y)$, so that when
brought together, due to their phase coherence, they produce the same
donut shape as the other states
(when~$\alpha_\to=i\alpha_\uparrow$
with~$|\alpha_\to|^2=1$). States with no coherence, like Fock or
thermal states, on the other hand, are indifferent to the choice of
basis, i.e., both
$\ket{1_\la}_\mathrm{S}\ket{1_\ra}_\mathrm{S}$ and
$\ket{1_\to}_\mathrm{S}\ket{1_\uparrow}_\mathrm{S}$ produce the donut
for both~$\mathrm{S}\in\{\mathrm{F},\mathrm{B}\}$.

\begin{figure}
  \includegraphics[width=\linewidth]{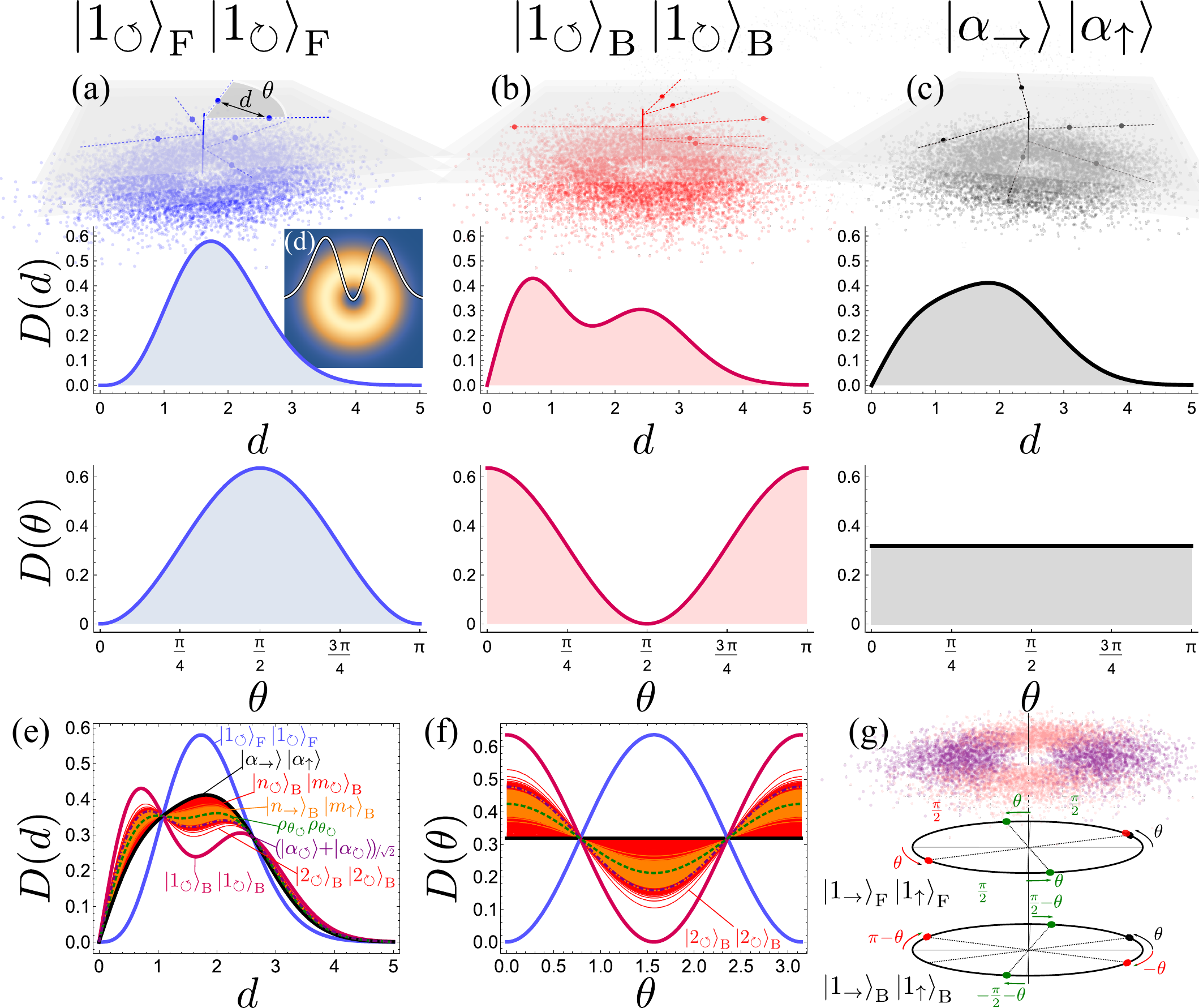}
  \caption{Top row: Single-shot realizations and their averages for
    Fock states~$\ket{1_\circlearrowleft}\ket{1_\circlearrowright}$
    with (a) Fermi~F \& (b) Bose~B statistics, and (c) for coherent
    states. The sum over many realizations converge to the exact same
    vortex shape shown in~(d).  Second row: distributions~$D(d)$
    between any two particles from the corresponding column. Bosons
    present a bimodal distribution. Third row:
    distributions~$D(\Delta\theta)$ of the angles between
    particles. (e--f) The three cases superimposed along with an
    enlarged family of quantum states, including unbalanced Fock
    states and thermal states. (g) Monte Carlo of distinguishable
    photons sampled from two dipoles, to reveal the underlying
    structure, and the angular correlations for fermions and bosons
    when these are indistinguishable in the donut. A particle at
    angle~$\theta$ (black) suppresses red and stimulate green ones,
    resulting in rich spatial correlations.}
\end{figure}

Despite this mathematical identity of their one-photon reduced density
matrix, all these states differ drastically for their two-photon
reduced density matrix, as should be clear from
Eqs.~(\ref{eq:Wed27Sep151754BST2023}). A more ``visible'' and direct
manifestation of such departures is shown in the figure, where we
compare the three illustrative cases of (a~\&~b) two Fock states with
one particle in each vortex state and (c) two coherent states with
mean amplitude one, so there is also one particle in each mode but
this time on average and with poissonian fluctuations (thus also
having no particle with the same probability~$\approx 37\%$ than
having one, and with more than one particle over one-fourth of the
time). Although the density profiles for all cases, as reconstructed
from averaging over the photons detected in isolation, are identical
and recover the theoretical limit (shown in the Inset~(d) both in~2D
and with a cut along the radius), photons detected in pairs behave
differently.  The simplest quantity to measure is the distance between
them. This is obtained from~$\rho^{(2)}$ as
$D(d) \equiv\langle{:}\hat N^2{:}\rangle^{-1}{\int
  \rho^{(2)}(\mathbf{r},\mathbf{r}') \delta(\|\mathbf{r}-\mathbf{r}'\|
  - d)\, \mathrm{d}\mathbf{r}\, \mathrm{d}\mathbf{r}'}$
with 
$\langle{:}\hat N^2{:}\rangle= \int
\rho^{(2)}(\mathbf{r},\mathbf{r}')\, \mathrm{d}\mathbf{r}\,
\mathrm{d}\mathbf{r}'$ the normalization. The average is, this time,
over (at least) two-photon observables. We show three such frames for
each case at the top of the figure. Superimposing all these frames,
one averages over the two-particle observables to recover the
one-particle one as
$\rho^{(1)}(\mathbf{r})=\int\rho^{(2)}(\mathbf{r},\mathbf{r}')\,\mathrm{d}\mathbf{r}'$. Experimentally,
this corresponds to the acquisition of the density profile by
integrating over the detected photons, washing out their correlations
and properties in the process. Multiphoton correlations must thus be
obtained from independent frames before their averaging, which
constitutes a quantum measurement.  The corresponding theoretical
distributions 
for the three cases are shown below, with a singled-peak
$D_\mathrm{F}(d)={1\over2}d^3\exp(-d^2/2)$ for Fermions, a bimodal
$D_\mathrm{B}(d)={1\over8}(8-4d^2+d^4)\exp(-d^2/2)$ for bosons and a
flattened $D_\mathrm{Coh}(d)={1\over16}d(8+d^4)\exp(-d^2/2)$ for
coherent states.  Although bosons, with a mean separation of
$\sqrt{121\pi/128}\approx 1.72$ time the vortex radius, are closer on
average than fermions, whose mean separation
is~$\sqrt{9\pi/8}\approx 1.88$ (all distances are in units of the
vortex radius), they present two local maxima, at both very small
distances of~$\approx0.71$ and large ones~$\approx 2.4$ as opposed to
$\sqrt{3}\approx 1.73$ for fermions. In fact, bosonic Fock states are
those most likely to distribute their two photons farther apart than a
distance strictly greater than~$d=2$. Curiously, all these quantum
states have the same probability ($3/e^2$) to be found at a distance
equal to the vortex diameter, and this remains true for the extended
family considered in the Supplementary~\cite{sup}. Also, all the
distance distributions feature the same variance
$\operatorname{Var}(d) = 4$ despite corresponding to different quantum
states. There are thus various peculiarities from even simple cases
put in various quantum states.  The bimodal curiosity is easily
explained, as due to the spatial profile of the underlying state (the
donut shape). If instead of the distances---which is easy to measure
as this is absolute---we consider relative angles~$\Delta\theta$ with
respect to the donut center, then we can capture the essence of the
phenomenon. This is shown in the third row, where we plot
$D(\Delta\theta)\equiv{1\over\langle{:}\hat
  N^2{:}\rangle}\int\rho^{(2)}(\mathbf{r},\mathbf{r}')\delta(\vartheta'
- \vartheta -
\Delta\theta)\,\mathrm{d}\mathbf{r}\mathrm{d}\mathbf{r}'$ for the
respective cases. 
This results in the simpler
$D_\mathrm{F}(\Delta\theta)={2\over\pi}\cos^2\Delta\theta$ and
$D_\mathrm{B}(\Delta\theta)={2\over\pi}\sin^2\Delta\theta$ while
$D_\mathrm{Coh}(\Delta\theta)=\mathrm{cste}$ ($1/\pi$ as normalization
of the uniform distribution). This means that the geometric aspect
which bosons and fermions extremize is not the distance, but the
angle: fermions want to be perpendicular while bosons want to be
aligned. The boson bimodal distribution is because if particles get
aligned on the same side of the donut, they are close together, while
if they are on the other side of the hole, they are farther apart than
fermions, which are at right angle. Symmetrizing the angle alone is
possible because vortices have a hole in their center, and fermions
can never sit on each other as their only meeting point would be at
the core, which has no particles. Aligned bosons can be found at the
same position, but they do not maximize their spatial overlap by being
on the same side of the donut. Their probability is still locally
twice as much to be found at close distance~$d\approx0$ since
$D_\mathrm{B}(d)\approx d$ while $D_\mathrm{Coh}(d)\approx d/2$
(compare with $D_\mathrm{F}(d)\approx d^3/2$) but this ``spatial
condensation'' is because the perpendicular options are depleted
(while the opposite one, maximizing distances, is also Bose
stimulated).  Coherent states wash out all traces of particle
correlations, although bosonic symmetry is knitted in the fabric of
the wavefunction, but this cancels out by disentangling, so that
particles get distributed indiscriminately even in individual
frames. The second row of the figure is thus best understood as the
distribution for the law of
cosines~$D\equiv\sqrt{R_1^2+R_2^2-2R_1R_2\cos(\Theta)}$ for the
independent random variables~$R_1$ and~$R_2$ sampling the radial
profile~$\rho_1(r)$ while~$\Theta$ follows~$D_\mathrm{F}$,
$D_\mathrm{B}$ or~$D_\mathrm{Coh}$, confirming that particles do not
``care'' about their distance per se, which is sampled randomly and
independently, but about their angle, wherein lies the two-particle
correlations.  There are countless other quantum states that can
correlate their photon pairs in a way which does not transpire through
single-photon observables, and the most straightforward extensions to
the Fock and coherent states are shown in Panels~(e) for distances
and~(f) for angles, enclosed between the Bose and Fermi extrema of two
particles. Cothermal states span continuously between the thermal and
coherent boundaries and a precise measurement of this distribution
would allow to extract information on the underlying quantum state,
such as the coherent fraction of a Bose condensate.  Vortices from two
particles in the dipole basis are interesting in this respect.
Fermions retain the same correlations in both bases, with the
tendency, when a photon is detected at an angle~$\theta$, to have the
second one be detected at angles~$\theta\pm{\pi\over2}$ while
suppressing those at~$\theta+\pi$, as sketched in panel~(g). This is
clear from the quantum state expressed in both basis:
$\ket{1_\to,1_\uparrow}_\mathrm{F}=i\ket{1_\la,1_\ra}_\mathrm{F}$.
Bosonic dipoles, on the other hand, turn into NOON vortices
$\ket{1_\to,1_\uparrow}_\mathrm{B}={i\over\sqrt2}(\ket{2_\la,0_\ra}-\ket{0_\la,2_\ra})$
with, surprisingly, the distribution of distances of uncorrelated
particles, i.e., $D_\mathrm{coh}(d)$. But this is not because they
are, like coherent states, not correlated. On the opposite, they have
even more complex correlations, requiring to keep track of two
angles. We can indeed write the two-particle reduced density matrix
for all cases in polar coordinates~as
\begin{equation}
  \label{eq:Thu18Jan220356CET2024}
  \rho^{(2)}(r,s,\theta,\vartheta)={2}r^2s^2e^{-r^2-s^2}D(\theta,\vartheta)/\pi^2\,,
\end{equation}
where, in most cases, the two-angle distributions is rotation
invariant and thus depend only on~$\Delta\theta\equiv\theta-\vartheta$
as is the case for the distributions already given and
plotted. Bosonic dipoles, on the other hand, have
$D_{\ket{1_\to,1_\uparrow}_\mathrm{B}}(\theta,\vartheta)={2\over\pi}\sin^2(\theta+\vartheta)$
with an absolute dependence of the angles (the rotation invariance is
broken by the dipole orientation). This is sketched on the figure with
a ``first'' detected photon (in black) at~$\theta$ finding others more
likely or suppressed, according to the bosonic stimulation of vortices
at diagonal angles ($\theta=k\pi/4$ for integer~$k$), but of fermionic
ones at right angles ($k\pi/2$). This corresponds to antibunching of
same-dipole arrangements and bunching of overlapping perpendicular
dipoles. The averages of these competing cases result in uncorrelated
distances, as if the resulting vortex was coherent, when it is, in
fact, richly structured.  Similarly, thermal states have the same
correlations in either basis by averaging unbalanced Fock
states~$\ket{n,m}$~\cite{sup}.

An exciting immediate follow-up of our approach is to query the
bosonic character of excitonic vortices from such spatial correlations
for two electron-hole pairs bound as excitons, so bosons, and thus
expressing as a whole bunching tendencies, but with a built-in
frustration of their constituents opposing Pauli
repulsion~\cite{laussy24a}.  Identifying which model, from simple
phenomenological ones based on $q$-deformed algebra to exact full
semiconductor Hamiltonians treatment, reproduces unequivocal
experimental results, would resolve this thorny and longstanding
controversy of solid-state
physics~\cite{laikhtman07a,combescot08a}. Then one can turn to a full
dynamical (not stationary), interacting (not linear), multiphoton
($>2$) correlations in both space and time for arbitrary spatial
profiles (not merely vortices). Two and three-photon vortex states
have recently been created in strongly-interacting Rydberg
gases~\cite{drori23a}, which also provide natural and even richer
platforms for such characterizations. There is no end of problems to
tackle in this way. Surprising correlations were noted early on from
sub-atomic physics with unexpected bosonic effects even between
distinguishable particles, due to intermediate-state and interference
effects in multipion production~\cite{andreev91a}. Interestingly, this
was linked to some innate capacity of the particles to produce
squeezing.  From our quantum-optical perspective, one can get direct
control of the quantum states of the field to cover the full gamut of
possibilities beyond the Bose-Einstein correlations found in stars and
nucleis, and thus go at the bottom of these fascinating effects which
started the modern theory of light.\\

\textbf{Acknowledgments:} This work was supported by the
HORIZON~EIC-2022-PATHFINDERCHALLENGES-01 HEISINGBERG project
101114978.

\bibliographystyle{naturemag}
\bibliography{sci,arXiv,books,spatial-correlations}

\end{document}